\documentclass[11pt,twoside]{article}

\usepackage{asp2004}
\usepackage{graphicx}
\usepackage{lscape}
\usepackage{sidecap}
\usepackage{multicol}

\begin{document}
\title{Dust in spiral disks:Ê
opacityÊprofiles from FIR emission and counts of distant galaxiesÊ}   
\author{B. W. Holwerda\altaffilmark{1}, R. A. Gonz\'{a}lez \altaffilmark{3}, D. Calzetti\altaffilmark{1}, R. J. Allen\altaffilmark{1}, P. C. van der Kruit \altaffilmark{2} and the SINGS team}

\altaffiltext{1}{Space Telescope Science Institute, Baltimore, MD 21218}
\altaffiltext{2}{Kapteyn Astronomical Institute, P.O. Box Groningen, the Netherlands.}
\altaffiltext{3}{Centro de Radiastronom\'{\i}a y Astrof\'{\i}sica, Universidad Nacional Aut\'{o}noma de M\'{e}xico, 58190 Morelia, Michoac\'{a}n, Mexico}

\begin{abstract} 

Dust emission in the far-infrared (FIR) characterizes the temperature and quantity of 
interstellar dust in a spiral disk.  The three Spitzer/MIPS bands are well suited to 
measuring the gradient in temperature and the total optical depth in the disk of a 
spiral galaxy. Another way to estimate the quantity of dust in a spiral disk is the 
"Synthetic Field Method" (SFM, \cite{Gonzalez98}), which uses the number of distant 
field galaxies seen through the disk of the nearby spiral.  The optical depth estimated 
from this method can be compared to the values derived from the FIR emission.  
Since the two techniques depend on different assumptions regarding the dust 
geometry and emissivity, this comparison between the optical depth profiles can 
potentially shed light on the structure and quantity of the ISM in spiral disks, especially any 
colder components. 
The dust responsible for the opacity from distant galaxy counts appears to be 
predominantly cold ($T \leq ~ $20 K.). The differences between the radial absorption 
profiles can be explained by spiral arms in the SFM measurements. Taken over the 
same aperture, galaxy counts show higher extinction values than the FIR derived ones.
The implications for dust geometry can hopefully be explored with 
a more rigorous estimate of dust mass from the FIR fluxes.
\end{abstract}

\section{Introduction}

The dust content of spiral disks can be measured from Far-Infrared (FIR) or sub-mm 
emission, reddening or by dust absorption of a known background source. 
The emission depends on the temperature and mass of the dust in the spiral disk. 
The number of distant galaxies depends on the average optical depth and hence 
cloud cover. By comparing these two measurements, we hope to probe the ISM 
geometry in spiral disks. 

\begin{figure}
  \begin{center}
    \begin{minipage}{0.55\linewidth}
      \includegraphics[width=1.2\textwidth]{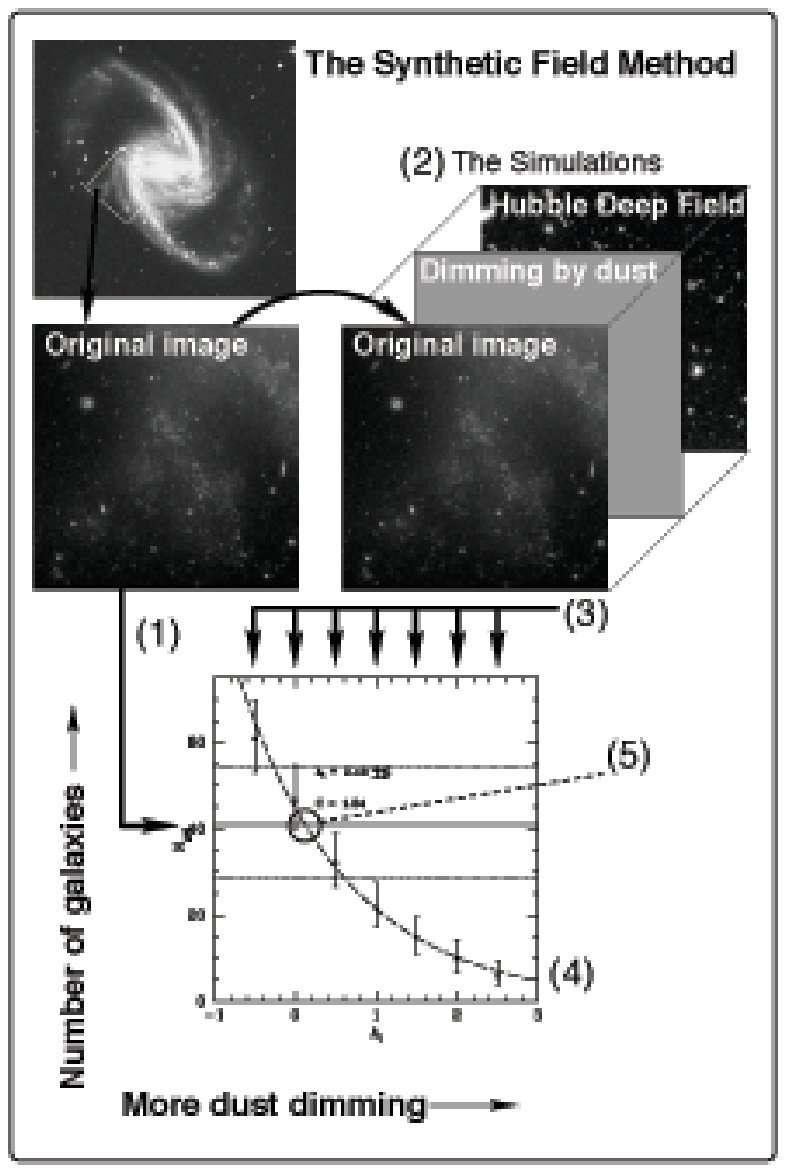}
    \end{minipage}\hfill
    \begin{minipage}{0.45\linewidth}
      \caption{A schematic of the ``Synthetic Field Method'' \citep{Gonzalez98, Holwerda05}. 
	First a WFPC2 field is retrieved from the Hubble Space Telescope archive and redrizzled. 
	The SFM steps are:\protect\\ 
	1. The number of distant galaxies in the original science field are counted. \protect\\ 
	2. The ``synthetic fields'' are made by combining a dimmed Hubble Deep Field 
	with the science field.\protect\\ 
	3. The numbers of synthetic galaxies are counted in the synthetic fields.\protect\\ 
	4. $\rm A = -2.5 ~ Log_{10}(N/N_0)$ it fitted to the number of synthetic galaxies ($N$) 
	as a function of the applied dimming ($A$).\protect\\ 
	5. From the intersection between the number galaxies in the science field and 
	the fit, the average dimming in the image is found. \label{fig:SFM}}
    \end{minipage}
  \end{center}
\end{figure}

The number of distant galaxies seen through the disk of a galaxy in an HST image 
needs to be calibrated for crowding and confusion with the ``Synthetic Field Method'' (SFM, Figure \ref{fig:SFM}).
The FIR emission is measuremed from Spitzer/MIPS data taken for the SINGS project 
\citep{sings}. The emission can be translated into an optical depth with an estimate for 
the emissivity and the temperature similar to \cite{Alton98b}. 
First we compare the radial profiles from the MIPS images to the SFM profiles from 
\cite{Holwerda05b} for 10 galaxies common to both samples. In addition, the 
WFPC2 field-of-view is used as an aperture on the MIPS images and the opacity 
measures for that section of the disk are compared directly.

\section{Radial Profiles}

\begin{figure}
  \begin{center}
    \begin{minipage}{0.6\linewidth}
\includegraphics[width=1.0 \textwidth]{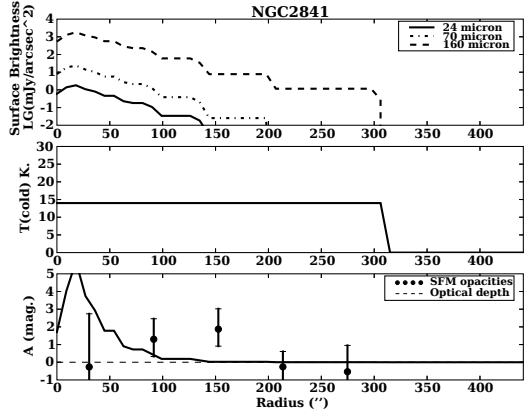}
    \end{minipage}\hfill
    \begin{minipage}{0.4\linewidth}
      \caption{The radial profiles of NGC2841. The MIPS profiles (top), the derived cold dust temperature profile (middle) and inferred opacity profile (bottom). The points are the opacity measurements from the counts of distant galaxies in HST data.\label{fig:n2841}}
    \end{minipage}
  \end{center}
\end{figure}

\begin{figure}
  \begin{center}
    \begin{minipage}{0.5\linewidth}
	\includegraphics[width=1.0\textwidth]{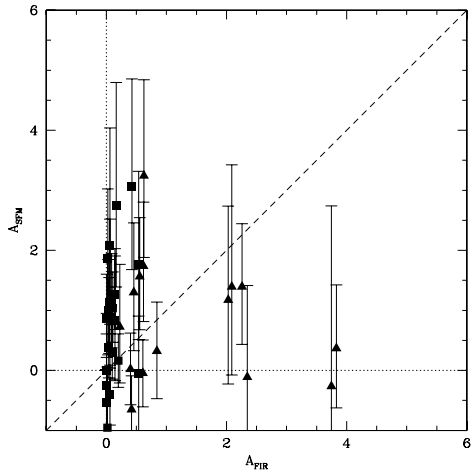}
	 \end{minipage}\hfill
    \begin{minipage}{0.5\linewidth}
	\caption{The absorption estimates from emission ($A_{FIR}$) and counts of distant galaxies ($A_{SFM}$) for 10 galaxies in our sample. The squares are for the outer radii 
($\rm R > 0.5 ~ R_{25}$) and the triangles for the inner parts of the disk ($\rm R < 0.5 ~ R_{25}$).\label{fig:tauA}}
   \end{minipage}
  \end{center}
\end{figure}

Figure \ref{fig:n2841} shows a typical profile. The FIR profiles are fitted with two thermal 
components ($T = 50 ~K$ and $T < 30 ~K$). The SFM profile is determined from the 
section of the disk covered by the WFPC2. In the case of NGC 2841, the bump at 150" 
 in the SFM profile appears due to the spiral arm. 
Figure \ref{fig:tauA} shows the FIR and SFM values from the radial profiles. The SFM 
values are higher for the outer disk while some of the inner disk show higher FIR 
derived opacities. This could be explained by the difference in field-of-view over 
which these radial profiles were measured. 


\section{WFPC2 Apertures}

\begin{figure}
  \begin{center}
    \begin{minipage}{0.5\linewidth}
	\includegraphics[width=1.0\textwidth]{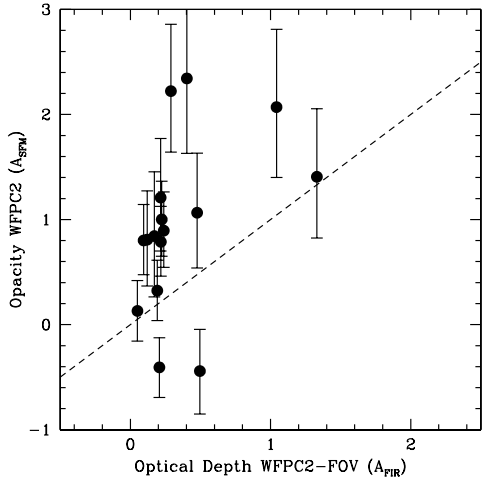}
	 \end{minipage}\hfill
    \begin{minipage}{0.5\linewidth}
	\caption{The absorption estimates from emission ($A_{FIR}$) and counts of distant galaxies ($A_{SFM}$) for the section of the disks covered by the WFPC2. Negative values from the counts of galaxies may very well be due to cosmic variance in those numbers, which is reflected in their uncertainties. \label{fig:wfpc2tauA}}
   \end{minipage}
  \end{center}
\end{figure}

To remedy the above problem with the radial profiles, the opacity and FIR flux within 
the WFPC2 FOV can also be compared. Figure \ref{fig:wfpc2tauA} shows the ratio of 
FIR derived opacity and the one based on counts of galaxies. The majority of the SFM 
opacities is still greater than the FIR ones, indicating a high filling factor for the cloud 
volume. The implication could be that (a) the average cloud size is very small, or (b) the clouds are predominantly oblate in the plane. 

\section{Conclusions}

\begin{itemize}
\itemsep=0.1pt
\item[1.] The dust found by the SFM  is predominantly cold ($T < 20 ~ K$.).
\item[2.] The higher SFM radial profile in the outer disk most likely due to spiral arms.
\item[3.] Compared over the same aperture, SFM values remain higher than the FIR ones. 
\end{itemize}
\vspace{-4pt}
Future work comparing FIR emission and galaxy counts should include an improved 
estimate of the dust mass from FIR, since a 2 temperature component fit is too simplistic. 

\footnotesize{
\acknowledgements This work is based on data obtained with the Spitzer and Hubble Space Telescopes, which are operated by JPL, CalTech and STScI respectively for NASA and AURA.
Support for this work was provided by NASA grant c3886 to D. Calzetti. 
\begin{multicols}{2}

\end{multicols}
}
\clearpage


\end{document}